\documentclass[12pt]{revtex4}
\usepackage{graphicx}
\begin{document}
\title{Testing Doppler type shift for an accelerated source and determination of the universal maximal acceleration}

\author{ Y. Friedman\\
Jerusalem College of Technology\\P.O.B. 16031 Jerusalem 91160, Israel\\
e-mail: friedman@jct.ac.il}

\begin{abstract}
An experiment for testing Doppler type shift for an accelerated source and determination of the universal maximal acceleration is proposed.
\end{abstract}
\maketitle
\section{Introduction}

Based on the generalized principle of relativity and the ensuing symmetry, we have shown \cite{F04}, \cite{FG4} and \cite{FConc} that there are only two possible types of transformations between uniformly accelerated systems. Using the proper velocity-time description of events rather than the usual space-time description, these transformations become linear. The first allowable type of transformation is Galilean and holds if and only if the Clock Hypothesis is true.

If the Clock Hypothesis is not true, it was shown that the transformation is of Lorentz-type and implies the existence of a universal maximal acceleration $a_m$. For this case it was shown \cite{F09} that a Doppler type shift for an accelerated source will be observed. This Doppler type shift is similar to the  Doppler shift due to the velocity of the source. The formulas for this shift are the same as those for the velocity Doppler shift, with $v/c$ replaced by $a/a_m$.

In W. K\"{u}ndig's \cite{Kundig}  measurement of the transverse Doppler shift, the source was accelerated and  exposed also to a longtitudal shift due to the acceleration. This experiment, as reanalyzed  by Kholmetskii et al, \cite{Khoimetski} shows that the Clock Hypothesis is not valid. Based on the results of this experiment, our theory predicts \cite{F09}  that the value of the maximal acceleration $a_m$ is of the order $10^{19}m/s^2$.

 \section{Proposed experiment}

 We propose an experiment which will be able to prove the existence of a longtitudal Doppler type shift for an accelerated source and will determine the exact value of the universal maximal acceleration.
 
In the experiment we plan to rotate a disk of radius about 5cm. On this disk we will mount a Mossbauer source $MS$ and an absorber $A$, diametrically opposed as in Fig 1.

\begin{figure}[h!]
  \centering
%\scalebox{0.4}{\includegraphics{pointChargePoten}}
\scalebox{0.8}{\includegraphics{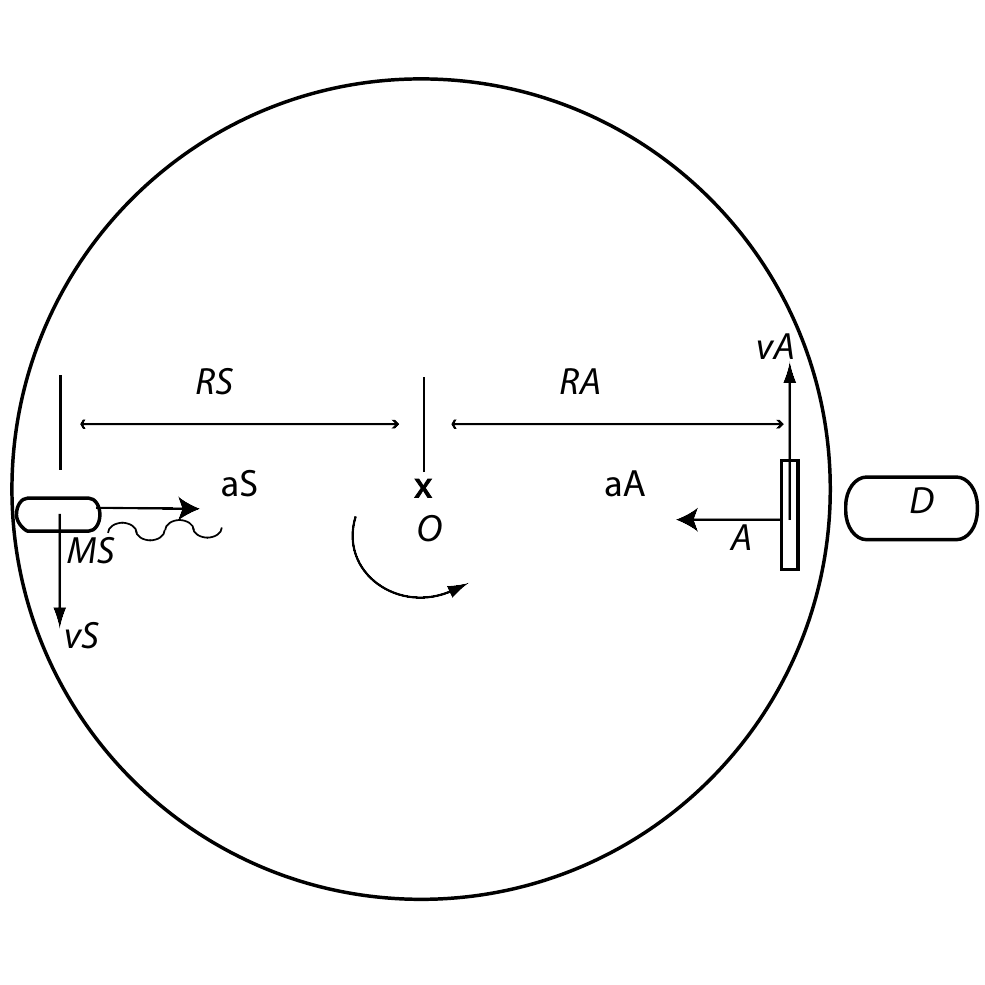}}
  \caption{The rotating disk of the proposed experiment.}\label{MaxAccelExp}
\end{figure}

  The detector $D$ will be outside the rotating disk. We denote the distance of the source $MS$ from the center $O$ of the disk by $RS$ and the distance of the absorber $A$ from the center by $RA$. The disk will be rotated by a vibrationless rotor with several angular velocities $\omega$ ranging up to 30,000 rpm or higher.

  The frequency of the radiation from the source $MS$ reaching the absorber $A$ undergoes two types of shifts, the first one due to  the relative time dilation between them, and the second one due to the acceleration between them. By varying  the position of the absorber $RA$, we want to find the point $R^*$ where there will be zero shift. This will be expressed by minimum count at the detector. At this distance $R^*$  of the absorber, the transversal velocity Doppler shift is canceled by the Doppler type shift due to acceleration.

\section{Experiment analysis}

The source and the absorber have velocities $vS=\omega RS$ and $vA=\omega RA$, respectively, where $\omega$
is the angular velocity of the rotor.  These velocities are transversal to the direction of radiation. Hence, the radiation undergoes a transversal Doppler shift (see \cite{Rindler}), based on the relative time dilation between the source and the absorber. Thus, the transversal Doppler shift will be
\begin{equation}\label{dopDif}
   \frac{\sqrt{1-\frac{\omega ^2 RA^2}{c^2}}}{\sqrt{ 1-\frac{\omega ^2 RS^2}{c^2}}}\approx
   \sqrt{1+\frac{\omega ^2 }{ c^2}(RS^2-RA^2)}=1+\frac{\omega ^2 }{2 c^2}\Delta R(RS+RA),
\end{equation}
where $\Delta R=RS-RA$.

The $MS$ source has also an acceleration $aS=\omega ^2 RS$ toward the center of the disk, while the absorber has an
acceleration $aA=\omega ^2 RA$ toward the center of the disk. Thus, the relative acceleration of the source toward the absorber will be
\begin{equation}\label{relat accel}
   a=aS+aA=\omega ^2 RS+\omega ^2 RA=\omega ^2 (RS+RA)\,.
\end{equation}
According to our theory, \cite{F09}, the longtitudal Doppler type shift due to the acceleration is
\[\left(1-\frac{a}{a_m}\right)=1-\frac{\omega ^2 (RS+RA)}{a_m}\,.\]

 The total shift between the source and the absorber will be
 \begin{equation}\label{totaltime dila}
    \left(1+\frac{\omega ^2 }{2 c^2}\Delta R(RS+RA)\right)\left(1-\frac{\omega ^2 (RS+RA)}{a_m}\right)\approx
    1+\omega ^2 (RS+RA)\left(\frac{\Delta R }{2 c^2}-\frac{1}{a_m}\right).
 \end{equation}
 If for some value $R^*$ of $R2$ we get zero shift, then for this $\Delta R^*=RS-R^*$ we get
 \[\frac{\Delta R^* }{2 c^2}-\frac{1}{a_m}=0,\]
 which establishes the value of the maximal acceleration as
 \begin{equation}\label{am formula}
  a_m=\frac{2c^2}{\Delta R^* }.
 \end{equation}

 Note that for $a_m=10^{19} n/s^2$, we expect $\Delta R^* \approx 2cm$.

We will start with $R2=R1$ and decrease the value of $R2$ until we get a minimum of the count of the detector.
From this minimal value, we can use (\ref{am formula}) to determine the value of the maximal acceleration.
\section{Discussion}
Other experiments to measure the size of transversal Doppler shift by use of Mossbauer effect were based on the measure of the
count at the detector. To be able to translate this count into Doppler shift, the count was compared with a similar count for a known longtitudal Doppler shift and assuming that the fast rotation of the system will not change significantly this count. Such assumption is very questionable.  By using a complex setup, in the K\"{u}ndig experiment the shift was measured without this assumption.
 
 In our experiment need only to define the position of the absorber were there is no shift, which do not need any assumptions on the connection between the count at the detector and the size of the shift. The unknown Doppler shift due to acceleration will be derived from the known transversal Doppler shift due to the time dilation of a moving object. The time dilation of a moving object was verified by several experiments, see for example \cite{Baily77}.

\end{document}